\documentclass[12pt]{article}
\usepackage{latexsym}
\usepackage{graphicx}
\usepackage{caption}
\usepackage{psfrag}
\usepackage{amsmath,amssymb}
\usepackage{dsfont}
\usepackage{lscape,multirow}
\newcommand{\lbl}[1]{\label{eq:#1}}

\newcommand{\be}{\begin{equation}}
\newcommand{\ee}{\end{equation}}
\newcommand{\bea}{\begin{eqnarray}}
\newcommand{\eea}{\end{eqnarray}}

\newcommand{\nn}{\nonumber}

\def\e{{\,\rm e}}
\def\Li{{\,\rm Li}}

\begin{document}

\begin{titlepage}
\begin{flushright}
UGFT-281/11\\
CAFPE-151/11
\end{flushright}

\begin{flushright}
\end{flushright}
\vspace*{1.5cm}
\begin{center}
{\Large \bf Analytic Reconstruction of heavy-quark\\ two-point  functions at $\mathcal{O}(\alpha_s^3)$}\\[2.0cm]

{\bf David Greynat} \\[.5cm]

Departamento de F\'{i}sica Te\'{o}rica\\ Universidad de Zaragoza,
Cl Pedro Cerbuna 12, E-50009 Zaragoza, Spain.\\[1cm]

{\bf Pere Masjuan} \\[.5cm]

Departamento de F\'{i}sica Te\'{o}rica y del Cosmos and CAFPE,
Universidad de Granada, E-18071 Granada, Spain\\[1cm]

{\bf Santiago Peris} \\[.5cm]

Grup de F{\'\i}sica Te{\`o}rica \\ Universitat Aut{\`o}noma de Barcelona, 08193 Barcelona, Spain.\\[1cm]

\end{center}

\vspace*{1.0cm}

\begin{abstract}

Using a method previously developed, based on the Mellin-Barnes transform, we reconstruct the two-point correlators in the vector, axial, scalar and pseudoscalar channels from the Taylor expansion at $q^2=0$, the threshold expansion at $q^2=4m^2$ and the OPE at $q^2\rightarrow -\infty$, where $m$ is the heavy quark mass. The reconstruction is analytic and systematic and is controlled by an error function which becomes smaller as more terms in those expansions are known.

\end{abstract}

\end{titlepage}

\section{Introduction}\lbl{int}
\setcounter{equation}{0}
\def\theequation{\arabic{section}.\arabic{equation}}

Vector, axial-vector, scalar and pseudoscalar two-point correlators of heavy quarks are very instrumental  functions for the determination of fundamental QCD parameters such as the strong coupling constant, $\alpha_s$, and the heavy quark masses, $m$. Presently, the input for these determinations comes both from the fact that QCD has been ``solved'' in $e^+e^-$ experiments  \cite{Kuhn:2007vp,Chetyrkin:2009fv,Kuhn:2010vx,Dehnadi:2011gc} and also from the fact that QCD may be ``solved'' numerically on the lattice\cite{Allison:2008xk}. Comparison between these two types of approaches results, of course, in a nontrivial test of the theoretical ideas involved and serves to assess the progress in our detailed understanding of QCD.

The fact that the quark is heavy brings about the welcome simplification that perturbation theory is a valid approximation. However, even in this case, and already at $\mathcal{O}(\alpha_s^2)$, present state-of-the-art perturbative calculations do not allow a complete knowledge of these two-point functions for all values of the momentum $q^2$. Instead of this, only expansions around specific values of $q^2$ have been possible to obtain. In Refs. \cite{Chetyrkin:1997mb}-\cite{Baikov:2009uw} we summarize the vast literature showing the titanic effort of several groups to compute the expansion of these two-point functions at low energy ($q^2=0$), at the production threshold ($q^2=4m^2$) and at high energy ($-q^2\rightarrow \infty$). In some particular example, even up to 30 terms have been calculated. Such is the case of the low-energy expansion of the vector two-point function at $\mathcal{O}(\alpha_s^2)$ in Ref. \cite{BCS,MMM}. However, more often than not, and certainly at $\mathcal{O}(\alpha_s^3)$, we are limited by the fact that just a few terms in each expansion have actually been computed. For $\mathcal{O}(\alpha_s^3)$, a nice summary of all the information about these three expansions, gathered up to date,  can be found in Refs. \cite{Hoang:2008qy,Kiyo}.

Given the limited amount of information, a full reconstruction of the corresponding two-point function can only be approximate. In the pioneering work in Refs. \cite{Broadhurst:1993mw}-\cite{Chetyrkin:1998ix}, Pad\'{e} Approximants \cite{Baker,MPPades} were proposed as a method to simultaneously resum the three individual expansions. However, a Pad\'{e} Approximant is a rational function of $q^2$ and, therefore, cannot reproduce either the logarithms of the OPE or the square roots of the threshold expansion. Consequently,  one must first subtract from the original Green's function all this nonanalytic behavior with the help of some simple, but physically motivated, functions. Even so, some of the Pad\'{e}s constructed turn out to contain unphysical features, and should be discarded by means of some type of prescription. This results in some degree of arbitrariness inherent to the method and, of course, in a corresponding systematic error,  as was pointed out in, e.g.  Ref. \cite{Hoang:2008qy}. A common practice then is to estimate this systematic error by some variation over the different more-or-less arbitrary choices made along the process of construction of the Pad\'{e}s.

Of course, whenever possible, checks have been made to assess the reliability of this Pad\'{e}-inspired method.  Although normally there seems to be a (perhaps surprisingly) high level of success, we think it is important that the results may be compared with those obtained by a completely different technique, such as the one which will be used in this work. As a matter of fact, Pad\'{e}s are known to be notoriously unpredictable approximants, with seemingly capricious properties of convergence. A notable exception to this rule is when the function to be approximated is a Stieltjes function, i.e one with a positive definite spectral function, for which there is a theorem assuring convergence of certain Pad\'{e} sequences \cite{Baker,MPPades}. Notice however that, although the complete vacuum polarization functions (\ref{poldef}) are Stieltjes, a fixed order in the $\alpha_s$ expansion need not be. In particular, as one can see in Figs. \ref{fig.spectral_functions} below, the $\mathcal{O}(\alpha_s^3)$ spectral functions are clearly not positive definite. This means that the convergence of the Pad\'{e}-inspired method is \emph{not} guaranteed. As a matter of fact, past examples in similar contexts do exist where the method produced certain results with an expected accuracy, and a comparison to the exact answer revealed that this accuracy had been   overestimated \cite{Melnikov}.

In Ref. \cite{Greynat} an alternative method has been presented.  Unlike the Pad\'{e} method, which is heavily numerical, the method in \cite{Greynat} is completely analytic. The approximation is also systematic (e.g. it obeys a simple counting rule) and expresses  the expansion of any vacuum polarization function as a unique combination of polylogarithms plus a known polynomial in the conformal variable $\omega$. This variable maps the cut complex plane in $q^2$ onto a unit disc. Furthermore, a parametrization of our ignorance about the first term not included in the approximation allows one to make an estimate of the systematic error incurred.

In this work, we will obtain results for the vector, axial-vector, scalar and pseudoscalar vacuum polarization functions at $\mathcal{O}(\alpha_s^3)$ \footnote{Some of these results for the vector channel were already obtained in Ref. \cite{Greynat}} and  make a detailed comparison with the corresponding results using the Pad\'{e}-inspired method, when available.

\section{Definitions}\lbl{def}
\setcounter{equation}{0}
\def\theequation{\arabic{section}.\arabic{equation}}

We define the generalized currents in the four channels: scalar ($s$), pseudoscalar ($p$), vector ($v$) and axial-vector ($a$) as
\begin{equation}\label{currentdef}
    s=\overline{\psi} \psi\quad ,\quad p=i\ \overline{\psi}\gamma_5 \psi \quad ,\quad v_{\mu}=\overline{\psi} \gamma_{\mu}\psi\quad ,\quad a_{\mu}=\overline{\psi} \gamma_{\mu}\gamma_5  \psi\qquad ,
\end{equation}
and the corresponding vacuum polarization functions as
\begin{eqnarray}\label{poldef}
q^2\ \Pi^{s,p}(q^2) &=& i \int\ d^4x\ \mathrm{e}^{iqx}\ \langle0| \mathrm{T}\begin{bmatrix}
 s(x) & \!\!\!\!\!s(0) \\  p(x) & \!\!\!\!\!p(0) \\ \end{bmatrix} |0\rangle  \hspace{1.5 cm} \nn \\
   \left(q_{\mu} q_{\nu}- q^2 g_{\mu\nu}\right)\  \Pi^{v,a}(q^2) + q_{\mu} q_{\nu} \ \Pi_{L}^{v,a}(q^2) &=& i  \int\ d^4x\ \mathrm{e}^{iqx}\ \langle 0| \mathrm{T}\begin{bmatrix} v_{\mu}(x) & \!\!\!\!\! v_{\nu}(0) \\  a_{\mu}(x) & \!\!\!\!\! a_{\nu}(0) \nn \\  \end{bmatrix} |0\rangle \ .\\  \end{eqnarray}
We will refer to this set of four polarization functions globally as $ \Pi^{\chi}$ where the index $\chi$ will run over the four channels, i.e.  $\chi=s,p,v,a$.

The longitudinal polarization function $ \Pi_{L}^{v,a}$ vanishes in the vector channel and, in the axial channel, is related to the pseudoscalar function through a Ward identity, so it will not be considered any further. Moreover, topologies in which massless particles occur as intermediate states have to be treated separately since they give rise to  different analytic properties of the polarization function in the complex plane. Leaving these contributions with massless cuts for future work, here we will concentrate on those perturbative Feynman diagrams which cause the polarization functions  $ \Pi^{\chi}$ to have a cut in the complex plane starting at $q^2=4 m^2$, where $m$ is the (pole) mass of the heavy quark considered.

\begin{figure}
\begin{center}
\includegraphics[width=6in]{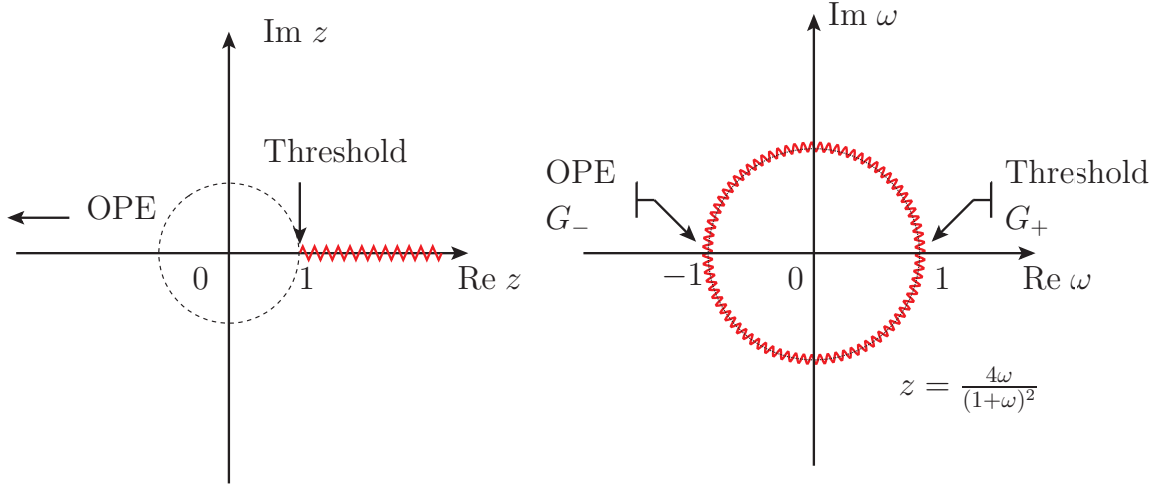}
\end{center}
\caption{Conformal mapping between $z$ and $\omega$, Eq. (\ref{conformal}). }
\label{omegaplan}
\end{figure}

Let us define the new variables $z= q^2/4m^2$ and $\omega$ as
\begin{equation}\label{conformal}
    z=\frac{4 \omega}{(1+\omega)^2}\qquad , \qquad \omega=\frac{1-\sqrt{1-z}}{1+\sqrt{1-z}}\  .
\end{equation}
This change of variables maps the cut $z$ plane into a unit disc in the $\omega$ plane, as we can see on Figure \ref{omegaplan}. The physical cut $z\in [1, \infty[$ is transformed into the circle $|\omega| = 1$ . The points $z = 0$ into $\omega = 0$, $z =1$ into $\omega=1$, the limit   $z \rightarrow +\infty \pm i \varepsilon$ into $\omega \rightarrow -1 \pm i \varepsilon$, and $z \rightarrow -\infty $ into $\omega \rightarrow -1$.

In terms of the variable $z$, the polarization function $\Pi^{\chi}(z)$ has the following properties:
\begin{enumerate}
    \item It admits a convergent expansion when $z\rightarrow 0$ since it is analytic in the disc $|z|<1$:
    \begin{equation}\label{eqT}
    \Pi^{\chi}(z)\underset{|z|<1}{=}\sum_{n=0}^{\infty}C^{\chi}(n)z^n\, ,
    \end{equation}

    \item It has a cut along $1\leq \mathrm{Re} (z) <\infty$. In the limit $z \rightarrow 1$, it can be expanded (threshold expansion) in the form
    \begin{equation}\label{eqA}
    \Pi^{\chi}(z)\underset{z\rightarrow 1}{\sim} \sum_{p,k}A^{\chi}(p,k)\  (1-z)^p \log^k(1-z)\, ,
    \end{equation}

    \item It admits and expansion (Operator Product Expansion) for $z\rightarrow - \infty$ such as
    \begin{equation}\label{eqB}
    \Pi^{\chi}(z)\underset{z\rightarrow - \infty}{\sim} \sum_{p,k}B^{\chi}(p,k)\ \frac{1}{z^p} \log^k(-4z)\, ,
    \end{equation}
\end{enumerate}

On the other hand, in terms of the variable $\omega$ the polarization function $\widehat{\Pi}^{\chi}(\omega)\equiv \Pi^{\chi}\big(4 \omega/(1+\omega)^2\big)$ obeys the following three expansions:
\begin{enumerate}
  \item The Taylor expansion ($\omega \rightarrow 0$):
  \begin{equation}
\widehat{\Pi}^{\chi}(\omega) \underset{|\omega|<1}{=}  \sum_{n= 0}^\infty \Omega^{\chi}(n)\; \omega^n \label{TaylorPi}\ .
\end{equation}
This expansion has radius of convergence $|\omega|<1$.

  \item The threshold expansion ($\omega \rightarrow +1$) and the Operator Product Expansion (OPE) ($\omega \rightarrow -1$) can be combined into:
      \begin{equation}
\widehat{\Pi}^{\chi}(\omega) \underset{\omega\rightarrow \pm 1}{\sim}\sum_{\lambda, p} \Omega^{\chi (\pm)}(\lambda, p)\; (1\mp \omega)^\lambda \log^p(1\mp \omega) \  , \label{ThresholdPiw}
\end{equation}
where the index $p$ is a (positive) integer but the index $\lambda$ may be integer (OPE), and also half-integer (threshold).
\end{enumerate}

The Taylor expansions of the function $\widehat{\Pi}^{\chi}(w)$, Eq. (\ref{TaylorPi}), and that of the function $\Pi^{\chi}(z)$, Eq. (\ref{eqT}),  are related. The relation between the two is given by
\begin{align}
\Omega^{\chi}(n) &= (-1)^n \sum_{p=1}^{n}  \frac{(-1)^p\;4^p\;\Gamma(n+p)}{\Gamma(2p)\Gamma(n+1-p)}\; C^{\chi}(p) \;,\label{Omega0=C}\\
 C^{\chi}(n)  &=\frac{\Gamma\left(n\right) \Gamma\left(\frac{1}{2}+n \right)}{\sqrt{\pi}}\;\sum_{p=1}^n \frac{ p}{\Gamma\left(1+n-p\right) \Gamma\left(1+n+p\right)} \;  \Omega^{\chi}(p)  \; .\label{C=Omega0}
\end{align}

It was shown in Ref. \cite{Greynat} that the asymptotic behavior of the $\Omega^{\chi}(n)$ for large $n$ is solely determined by the coefficients $\Omega^{\chi (\pm)}(\lambda,p)$ in Eq. (\ref{ThresholdPiw}) encoding non-analytic behavior, i.e. when $\lambda<0, \forall p$, or when $\lambda \geq 0, p\neq 0$. Furthermore, this asymptotic behavior is very precocious, yielding accurate results already for $n\gtrsim 2,3$. The combination of these two features allows the reconstruction of the vacuum polarization function  by carrying out the sum in Eq. (\ref{TaylorPi}) using the exact value for the first few coefficients $\Omega^{\chi}(n)$ up to a given value, say $n=N^*$, and then using the asymptotic expansion for $n>N^*$.

More specifically, the coefficient $\Omega^{\chi}$ admits the following expansion
\begin{equation}
\Omega^{\chi}(n)\underset{n\rightarrow \infty}{\sim} \Omega^{\chi}_{AS}(n) = \sum_{p,k} \left[ \alpha^{\chi}_{p,k}  + (-1)^n \beta^{\chi}_{p,k}\right]\frac{\log^k n }{n^p}\; , \label{DefOmega}
\end{equation}
where the coefficients $\alpha^{\chi}_{p,k}$ and  $\beta^{\chi}_{p,k}$ are in one-to-one correspondence with the coefficients $\Omega^{\chi (+)}(\lambda,p)$ and $\Omega^{\chi (-)}(\lambda,p)$ (respectively) in a systematic way. The precise connection is best obtained through: i) the Mellin-Barnes representation
\begin{equation}
\widehat{\Pi}^{\chi}(\pm\e^{-t}) \underset{t>0}{=} \int \limits_{c- i \infty}^{c+ i  \infty} \! \frac{ds}{2i\pi} \; t^{-s} \; \Gamma(s) \; \sum_{n=1}^\infty \; (\pm 1)^n\; \Omega^{\chi}(n)\;  n^{-s} \label{MellinPit}\; ,
\end{equation}
where $c$ is a constant located in the region of analyticity of the integrand in the complex $s$ plane, also known as the ``fundamental strip''; and ii) the Converse Mapping theorem dictionary, by which a function, $f(t)$,  and its corresponding Mellin transform, $\mathcal{M}[f](s)$, are related through
\begin{align}
\mathcal{M}[f](s) &= \int_0^\infty\! dt \; t^{s-1} \; f(t) & \rightleftharpoons  && f(t) &= \int \limits_{c- i  \infty}^{c+ i  \infty} \! \frac{ds}{2i\pi} \; t^{-s} \; \mathcal{M}[f](s) \label{dict1}\\
\mathcal{M}[f](s) & \asymp \sum_{p,k} \frac{r_{p,k}}{(s+p)^k} & \rightleftharpoons  && f(t) & \underset{t\rightarrow0}{\sim} \sum_{p,k} \frac{(-1)^{k-1}}{(k-1)!}\; r_{p,k}\;  t^p \log^{k-1} t \; , \label{dict2}
\end{align}
where the symbol $\asymp $ means that one sums over all the negative powers of the Laurent expansions of the function around every pole (i.e. the ``principal part'' of the meromorphic function). Feeding Eq. (\ref{MellinPit}) with the expansion (\ref{DefOmega}) and matching onto the expansion for $\widehat{\Pi}^{\chi}(\pm\e^{-t})$ on the lefthand side as given by (\ref{ThresholdPiw}) with the help of the dictionary (\ref{dict1},\ref{dict2}), one obtains all the coefficients $\alpha_{p,k}$ and  $\beta_{p,k}$ in terms of the threshold and OPE coefficients $A^{\chi}(p,k)$ and $B^{\chi}(p,k)$ in Eqs. (\ref{eqA},\ref{eqB})  in a completely systematic way, order by order in the expansions. For instance, if the leading term in the threshold expansion is given by the coefficient $A(-1, 0)$ in Eq. (\ref{eqA}) (as indeed it happens for $\chi=v,p$, see the third row in Table \ref{table-z}), one then obtains that the leading asymptotic behavior is given by (\ref{DefOmega}) with $ \alpha_{-1,0}= 4\ A(-1, 0)$ (see the corresponding entry in Table \ref{table-ab}). The relation between the coefficients $A$ and $B$ in the OPE and threshold expansion  (\ref{eqA},\ref{eqB}) and the coefficients $\alpha, \beta$ in the asymptotic expansion (\ref{DefOmega}) is exact. Although we will give numerical results for the sake of brevity, it is clear that the whole approach is fully analytic.

\section{ Results}\lbl{results}
\setcounter{equation}{0}
\def\theequation{\arabic{section}.\arabic{equation}}

All the relevant expansions have been conveniently listed in Refs. \cite{Hoang:2008qy,Kiyo} for the four polarization functions $\Pi^{\chi}$ with $\chi=v,a,p,s$. Their corresponding coefficients for the threshold expansion (\ref{eqA}) and the OPE (\ref{eqB}) can be extracted from these references and we collect them in Table \ref{table-z} (where $n_l$ stands for the number of light flavors). Alternatively, we can also express the four polarization functions in terms of the variable $\omega$ in Eq. (\ref{conformal}). After expanding, the corresponding coefficients for the OPE and threshold expansion  (\ref{ThresholdPiw}) can also be extracted and we collect them in Table \ref{table-w}.

\begin{table}
  \centering
{\footnotesize
\begin{tabular}{|c||c|c||c|c||c|c||c|c|}
\hline
 & \multicolumn{2}{c||}{$v$} & \multicolumn{2}{c||}{$a$} &\multicolumn{2}{c||}{$p$} & \multicolumn{2}{c|}{$s$}\\
\cline{2-9}
			$\Pi^\chi$					&	$n_l=3$ & $n_l=4$ 				&  $n_l=3$ & $n_l=4$ 			&	$n_l=3$ & $n_l=4$ 			 &	$n_l=3$ & $n_l=4$ \\
\hline
\hline
$A$(-1,0)   & 2.6364   & 2.6364   & 0  			& 0 		& 2.6364	& 2.6364  & 0			&	0			 \\
$A$(-1/2,0) &-25.2331  & -24.5549 & 0  			& 0 		& -22.9364  & -22.2582 & 0			&			 0			 \\
$A$(-1/2,1) & -7.7516  & -7.1774 & 0  			& 0 		& -7.7516   & -7.1774 & 0			&			 0			\\
$A$(0,1)    & -11.0654 & -9.1783 & -0.7311     &   -0.7311	& -3.8928   & -1.8142 & -1.0966 	& -1.0966	 \\
$A(0,2)$    & 1.4283   & 1.4732  & 0 			& 0 		& 3.2655    & 3.2826  &  0			&			 0			\\
$A(0,3)$    & -0.4219  & -0.3617 & 0 			& 0 		& -0.4219	& -0.3617 & 0			&			 0			 \\
\hline
$B(0,1)$   & -0.0699  & 0.0173   & -0.0699 & 0.0173   & 1.6431    &  1.5124   &  1.6431     &  1.5124 \\
$B(0,2)$   & 0.1211   & 0.0990   & 0.1211  & 0.0990   & 4.3309    & 3.7712    & 4.3309	   & 3.7712 \\
$B(0,3)$   & -0.0366  & -0.0310  & -0.0366 & -0.0310  & -1.1841   &-1.0668    &-1.1841	   & -1.0668 \\
$B(0,4)$   & 0 	      & 0 	     & 0 	   & 0		  & 0.0797    & 0.0723    & 0.0797      & 0.0723\\
$B$(1,1)  &  -3.7567  & -3.1062  & -3.3602 & -2.7103  &  -12.0583 & -10.3098  & -18.0487 	& -15.3645 \\
$B$(1,2)  &  2.1173   & 1.8983   & -2.5325 & -2.1623  &   3.0251  &  2.8886  & -5.5011     & -4.6759  \\
$B$(1,3)  & -0.3189   & -0.2894  & 0.8652  & 0.7774   & 0.4023    & 0.3532    &	3.7238		& 3.4077 \\
$B$(1,4)  & 0         & 0        & -0.0797 &-0.0723   & -0.1573   &-0.1467    &-0.4719		& -0.4402 \\
$B$(2,1)  & -5.1301   & -4.3252  & 4.5072  & 3.7977   & -5.8848   &-5.3034    & 11.8587 	& 10.3358 \\
$B$(2,2)  & 0.3182    & 0.3487   & -0.5490 & -0.5518  &  -1.1566  & -0.9612   &	 -4.5366 	& -4.3615 \\
$B$(2,3)  & 0.4015    & 0.3627   & -0.3190 & -0.2806  & 1.3083    &  1.2114   &	-1.1407 	& -1.0030  \\
$B$(2,4)  & -0.0787   & -0.0734  & 0.0787  & 0.0734   &  -0.1565  & -0.1475   &  0.4695     & 0.4426  \\
\hline
\end{tabular}
}
\caption{Threshold expansion and OPE coefficients in Eqs. (\ref{eqA},\ref{eqB}) for $n_l=3,4$; where $n_l$ stands for the number of light flavors.}\label{table-z}
\end{table}

\begin{table}
  \centering
{\footnotesize
\begin{tabular}{|c||c|c||c|c||c|c||c|c|}
\hline
 & \multicolumn{2}{c||}{$v$} & \multicolumn{2}{c||}{$a$} &\multicolumn{2}{c||}{$p$} & \multicolumn{2}{c|}{$s$}\\
\cline{2-9}
					$\Omega^\chi$			&	$n_l=3$ & $n_l=4$ 				&  $n_l=3$ & $n_l=4$ 			&	 $n_l=3$ & $n_l=4$ 			 &	 $n_l=3$ & $n_l=4$ \\
\hline
\hline
$\Omega^{\chi(+)}$(-2,0) & 10.5456  & 10.5456  & 0  	   & 0 			& 10.5456	& 10.5456 & 0		 &	0			 \\
$\Omega^{\chi(+)}$(-1,0) & 18.4286  &  18.6642 & 0  	   & 0 			&  13.8352  &  14.0708& 0		 &	0			 \\
$\Omega^{\chi(+)}$(-1,1) &  31.0063 &  28.7095 & 0  	   & 0 			&  31.0063  &  28.7095& 0		 &	0			 \\
$\Omega^{\chi(+)}(0,1)$  & -50.4189 & -45.0511 &  -1.4622  & -1.4622   	& -46.2611  & -40.3563&  2.1932  &  2.1932 	 \\
$\Omega^{\chi(+)}(0,2)$  & 12.7314  & 11.9097  & 0 		   & 0 			& 20.0801   & 19.1472 &  0		 &	0			 \\
$\Omega^{\chi(+)}(0,3)$  & -3.3750  & -2.8935  & 0 		   & 0 			& -3.3750	& -2.8935 & 0		 &	0			 \\
\hline
$\Omega^{\chi(-)}(0,1)$  & 0.4872   & 0.2979   & 0.4872    & 0.2979  	& -10.2951 & -7.9785  & -10.2951 & -7.9785	 \\
$\Omega^{\chi(-)}(0,2)$  & -0.7349  & -0.6359  & -0.7349   & -0.6359  	& -7.3644  & -7.0625  & -7.3644  & -7.0625	\\
$\Omega^{\chi(-)}(0,3)$  & 0.2932   & 0.2481   & 0.2932    & 0.2481     &  2.3993  & 2.1162   & 2.3993   & 2.1162		 \\
$\Omega^{\chi(-)}(0,4)$  & 0 		& 0 	   & 0 		   & 0			& 1.2757   & 1.1575   & 1.2757   & 1.1575		 \\
$\Omega^{\chi(-)}(2,1)$  &  0.1671  &  0.2416  & -2.2712   & -1.6012 	&  -1.5908 &  -1.2726 &  -3.3365 & -2.0631 \\
$\Omega^{\chi(-)}(2,2)$  &  0.7552  &  0.6947  & -0.7669   & -0.7815    &  4.5968  &  4.2651  &  0.0063  & -0.0398 \\
$\Omega^{\chi(-)}(2,3)$  &  -0.6378 &  -0.5787 & -0.0380   & -0.0497    &  -1.4088 &  -1.3911 & -1.7440  &-1.7921 	 \\
$\Omega^{\chi(-)}(2,4)$  & 0        & 0        & 0.3189    & 0.2894	    & 0.6292   & 0.5870   &	1.8876	 & 1.7610		 \\
$\Omega^{\chi(-)}(4,1)$  & 0.7588   & 0.7491   & -3.7157   & -3.0118    & -0.1969  & 0.1395   & -8.5663  & -6.9123  \\
$\Omega^{\chi(-)}(4,2)$  & -0.7823  & -0.7052  & -0.3778   & -0.4278    &  -0.9318 & -0.9492  & -3.1081  & -3.0457 		 \\
$\Omega^{\chi(-)}(4,3)$  & -0.4024  & -0.3532  & 0.6421    & 0.5518 	& 0.0547   & 0.0036   & 1.2480   & 0.9595 	 \\
$\Omega^{\chi(-)}(4,4)$  & -0.0787  & -0.0734  & 0.3976    & 0.3627 	&  0.4727  & 0.4395   & 2.3571   &  2.2036 \\
\hline
\end{tabular}
}
\caption{Threshold expansion and OPE coefficients in Eqs. (\ref{ThresholdPiw}) for $n_l=3,4$ (number of light flavors). }\label{table-w}
\end{table}

\begin{figure}
\centering
\includegraphics[width=3in]{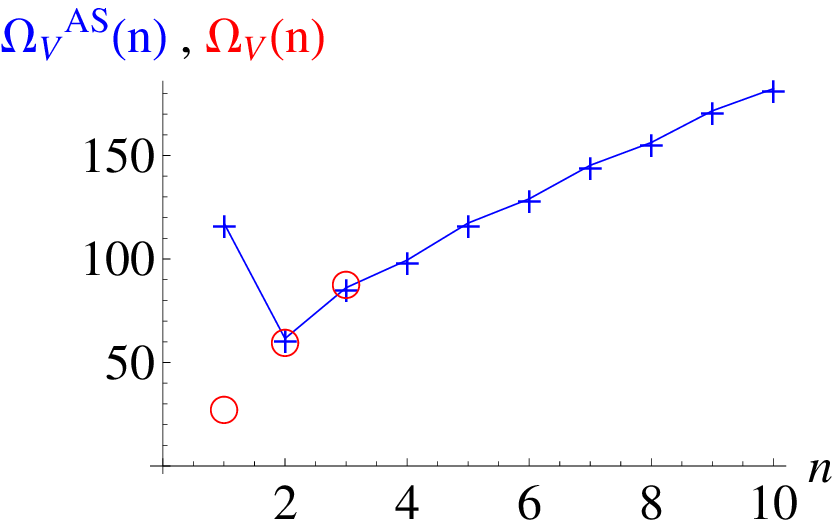}
\includegraphics[width=2.5in]{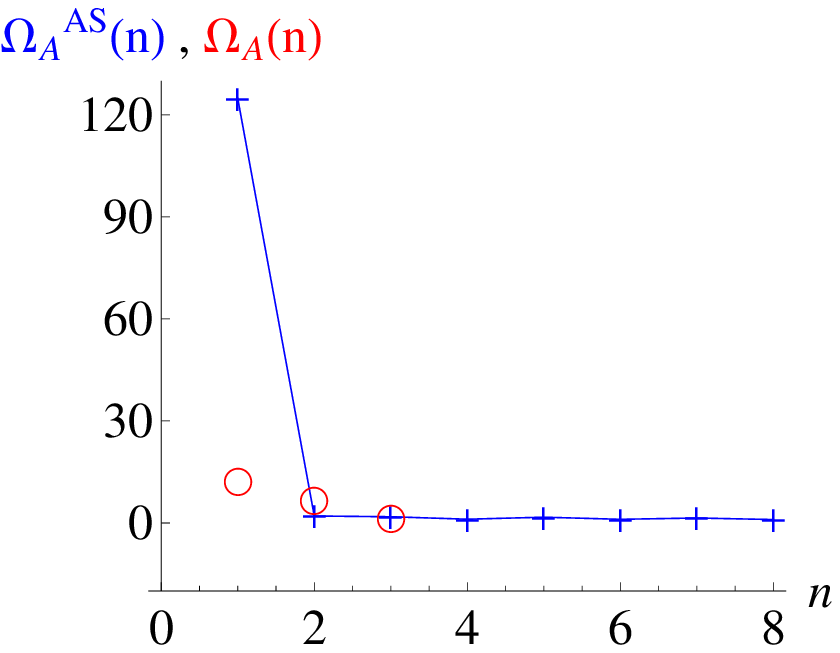}

\vspace{.5cm}

\includegraphics[width=2.7in]{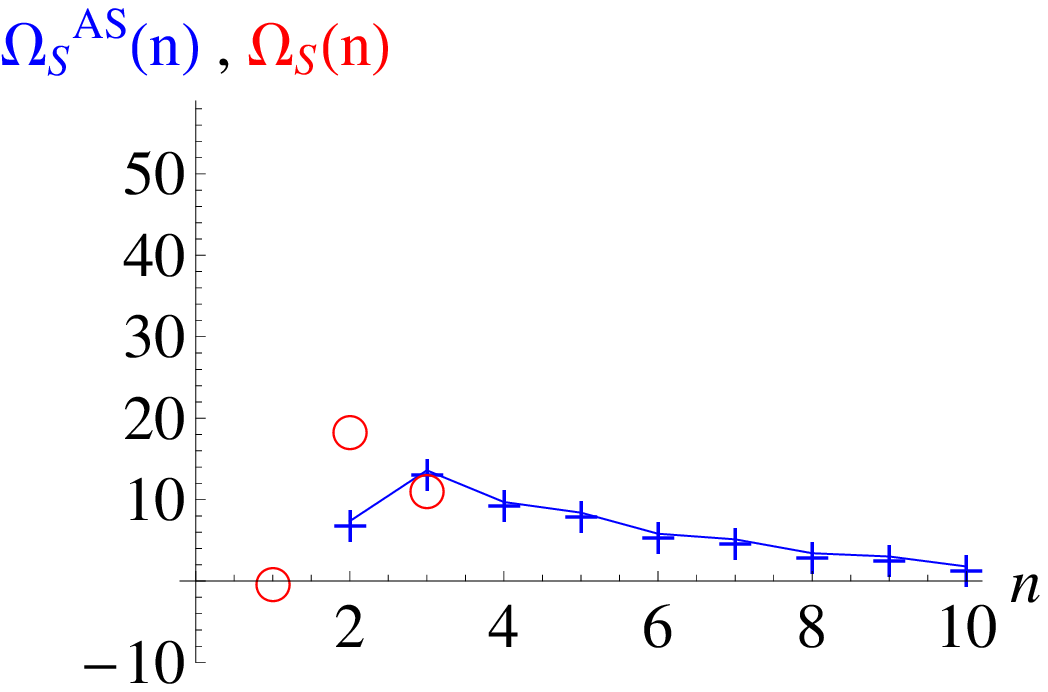}
\includegraphics[width=2.8in]{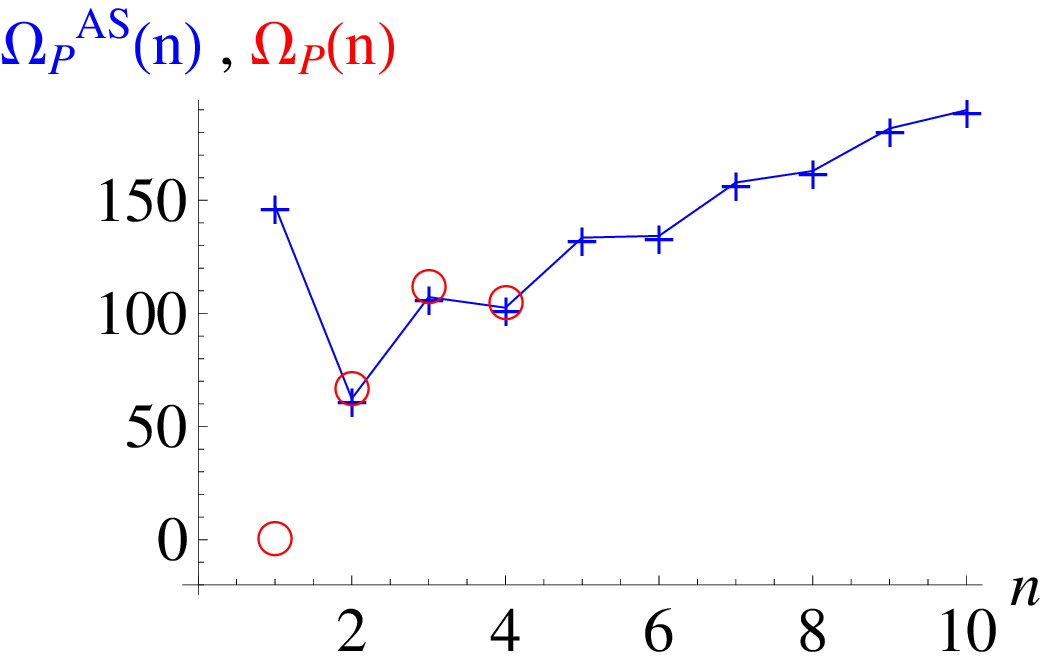}
\caption{Comparison of the asymptotic expression $\Omega_{AS}^{\chi}(n)$ (blue crosses) and the exact values  $\Omega_{AS}^{\chi}(n)$ (red circles), for $\chi=s,p,v,a$, extracted from Ref. \cite{Hoang:2008qy,Kiyo}. }\label{XXX}
\end{figure}

Using the matching condition (\ref{MellinPit}), one can deduce the values for the coefficients $ \alpha^{\chi}_{p,k}$ and $ \beta^\chi_{p,k}$ in the asymptotic expansion (\ref{DefOmega}). We list them in Table \ref{table-ab}. This asymptotic expansion can then be compared to the exact values of the first coefficients $\Omega^{\chi}(n)$ in Eq. (\ref{Omega0=C}), where the $C^{\chi}(n)$ are known exactly by means of diagrammatic methods. This is done in Figures \ref{XXX}. As one can see, the asymptotic expansion starts being a good approximation already for $n\gtrsim 2,3$. Note, in particular, the pseudoscalar channel where this comparison may be done up to $n=4$.

\begin{table}
  \centering
{\footnotesize
\begin{tabular}{|c||c|c||c|c||c|c||c|c|}
\hline
 & \multicolumn{2}{c||}{$v$} & \multicolumn{2}{c||}{$a$} &\multicolumn{2}{c||}{$p$} & \multicolumn{2}{c|}{$s$}\\
\cline{2-9}
				$\Omega^\chi_{AS}$				&	$n_l=3$ & $n_l=4$ 				&  $n_l=3$ & $n_l=4$ 			 &	$n_l=3$ & $n_l=4$ 			 &	 $n_l=3$ & $n_l=4$ \\
\hline
\hline
$\alpha^{\chi}_{{\text -1},0}$& 10.5456  & 10.5456	& 0	& 0 		& 10.5456	& 10.5456	& 0			 & 0 			 \\
$\alpha^{\chi}_{0,0}$ & -11.0769 & -12.6382 & 0			& 0 		& -6.4835   &-8.0448	& 0			& 0				 \\
$\alpha^{\chi}_{0,1}$ & 31.0063  & 28.7095	& 0			& 0 		& 31.0063	&28.7095	& 0		    & 0				 \\
$\alpha^{\chi}_{1,0}$ & 36.3318  & 33.0585	& 1.4622    & 1.4622    & 40.6575	&36.7189	& 2.1932	& 2.1932	 \\
$\alpha^{\chi}_{1,1}$ & 37.1514  & 33.8404	& 0			& 0 		& 51.8488	& 48.3155   & 0			& 0				 \\
$\alpha^{\chi}_{1,2}$ & 10.1250  & 8.6805	& 0 		& 0 		& 10.1250 	& 8.6805	& 0			& 0				 \\
\hline
$\beta^{\chi}_{1,0}$ & -0.1819   &-0.0555   &-0.1819    &-0.0555    & 9.9493    & 6.9861	& 9.9493	& 6.9861	\\
$\beta^{\chi}_{1,1}$ & -2.4852	 &-2.1312   & -2.4852   &-2.1312    & -43.1187  &-39.6735	& -43.1186  & -39.6735  \\
$\beta^{\chi}_{1,2}$ & -0.8795   &-0.7444   &-0.8795    &-0.7444    & 1.6381    & 1.6688	& 1.6381	& 1.6688	\\
$\beta^{\chi}_{1,3}$ & 0 		 & 0		&0			&	0		& 5.1027    & 4.6298	& 5.1027    & 4.6298	\\
$\beta^{\chi}_{3,0}$ & -10.4385  &-9.7282	& 26.2458	&22.9826    & 3.1298 	& 1.1687	& 93.7790	& 83.0590	\\
$\beta^{\chi}_{3,1}$ & -4.7750   &-4.2501	& -19.8617  &-18.4878   & -53.4944  &-50.0465	& -137.2835 & -129.2810 \\
$\beta^{\chi}_{3,2}$ & 3.8270    &3.4724	& -6.8349   &-6.1103    & 0.8960    & 1.1335	& -24.9630	& -22.4605  \\
$\beta^{\chi}_{3,3}$ & 0  		 &0			& 2.5513    &2.3149     & 5.0337    & 4.6960	& 15.1011	& 14.0879	\\
$\beta^{\chi}_{5,0}$ & -70.9277  &-63.8573	& 100.2171	&89.1103    & -115.8498	& -108.3750 & 440.1394	& 399.7520	\\
$\beta^{\chi}_{5,1}$ & 56.3093   &53.6862	& -72.4918  &-68.9185   & 62.1988	& 60.2675	& -512.9781 & -487.4843 \\
$\beta^{\chi}_{5,2}$ & 20.9951   &19.0619	& -29.3263  &-26.2676   & 38.4395   & 35.8466	& -129.9058 & -118.3556 \\
$\beta^{\chi}_{5,3}$ & -7.5506   &-7.0439	& 10.1019   &9.3589     & -9.9903   & -9.4668   & 60.1732	& 56.5763	\\
\hline
\end{tabular}
}
\caption{Values for the coefficients in the asymptotic expansion (\ref{DefOmega}) for $n_l=3,4$.}\label{table-ab}
\end{table}

Once the coefficients $ \alpha_{p,k}$ and $ \beta_{p,k}$ are known, one may reconstruct the vacuum polarization function $\widehat{\Pi}^{\chi}(\omega)$ by directly carrying out the sum in Eq. (\ref{TaylorPi}), splitting it in three different terms:
\begin{equation}
\label{ApproxPi2v1}
\widehat{\Pi}^{\chi}(\omega) =\sum_{n=1} ^\infty \Omega^{\chi}_{AS}(n)\ \omega^n +\sum_{n=1} ^{N^*} \left[\Omega^{\chi}(n) - \Omega^{\chi}_{AS}(n)\right]\omega^n + \mathcal{E}^{\chi}(N^*,\omega)  \;,
\end{equation}
where
\begin{equation}\label{errorfunction}
\mathcal{E}^{\chi}(N^*,\omega) \equiv \sum_{n=N^*+1} ^\infty \left[\Omega^{\chi}(n) - \Omega^{\chi}_{AS}(n)\right]\omega^n \;.
\end{equation}
Notice that the first $N^*$ coefficients, which are known from the Taylor expansion (\ref{TaylorPi}), have been included exactly, using the asymptotic expansion $\Omega^{\chi}_{AS}(n)$ only from the value $N^*+1$ onwards. The function $\mathcal{E}^{\chi}$ is an ``error function'' that will have to be estimated. As the expression (\ref{errorfunction}) clearly shows, $\mathcal{E}^{\chi}(N^*,\omega)$ becomes smaller as $N^*$ grows, for any $|\omega|<1$. One may also expect the error function to decrease as more terms in the series (\ref{DefOmega}) are included. This means that our approximation will become better as more terms from the three expansions (\ref{eqT}-\ref{eqB}) are known, as we explicitly show in the Appendix for the pseudoscalar channel.

\begin{table}
  \centering
{\footnotesize
\begin{tabular}{|c||c|c|c|c|}
\hline
$\chi$ & \multicolumn{2}{c|}{$\left[\Omega^\chi(n) - \Omega^\chi_{AS}(n) \right]_{n>N^*}$} & $N^*$ \\
\hline
\hline
\multirow{2}{*}{$v$}& $n_l=3$ 	& $\displaystyle \pm 15 \frac{\log^3 n}{n^2} \pm (-1)^n \mathcal{O}\left(\frac{\log^\ell n}{n^7}\right)$	& 3\\[1ex]
\cline{2-4}
										& $n_l=4$ 	&$\displaystyle \pm 15 \frac{\log^3 n}{n^2} \pm (-1)^n \mathcal{O}\left(\frac{\log^\ell n}{n^7}\right)$ 	& 3\\[1ex]
\hline
\hline
\multirow{2}{*}{$a$}& $n_l=3$ 	&$\displaystyle \pm 1 \frac{\log^3 n}{n^2} \pm (-1)^n \mathcal{O}\left(\frac{\log^\ell n}{n^7}\right)$		& 3\\[1ex]
\cline{2-4}
    								& $n_l=4$ 	&$\displaystyle \pm 1 \frac{\log^3 n}{n^2} \pm (-1)^n \mathcal{O}\left(\frac{\log^\ell n}{n^7}\right)$ 	& 3\\[1ex]
\hline
\hline
\multirow{2}{*}{$p$}& $n_l=3$		&$\displaystyle \pm 40 \frac{\log^3 n}{n^2} \pm (-1)^n \mathcal{O}\left(\frac{\log^\ell n}{n^7}\right)$ 	& 4\\[1ex]
\cline{2-4}
										& $n_l=4$ 	&$\displaystyle \pm 40 \frac{\log^3 n}{n^2} \pm (-1)^n \mathcal{O}\left(\frac{\log^\ell n}{n^7}\right)$ 	& 4\\[1ex]
\hline
\hline
\multirow{2}{*}{$s$}& $n_l=3$		&$\displaystyle \pm 10 \frac{\log^3 n}{n^2} \pm (-1)^n \mathcal{O}\left(\frac{\log^\ell n}{n^7}\right)$ 	& 3\\[1ex]
\cline{2-4}
										& $n_l=4$ 	&$\displaystyle \pm 10 \frac{\log^3 n}{n^2} \pm (-1)^n \mathcal{O}\left(\frac{\log^\ell n}{n^7}\right)$ 	& 3\\[1ex]
\hline
\end{tabular}
}
\caption{Error functions.}\label{totalerrors}
\end{table}

In this way, we could obtain the following approximation to the vector vacuum polarization function \cite{Greynat}:
\begin{align}\label{result3omega}
\widehat{\Pi}^{v}(\omega) & = \alpha_{\,-1,0} \; \frac{\omega}{(1-\omega)^2} + \alpha_{0,0} \; \frac{\omega}{1-\omega} - \alpha_{0,1} \Li^{(1)}(0,\omega) - \alpha_{1,0} \log(1-\omega) \nonumber\\
&- \alpha_{1,1}\Li^{(1)}(1,\omega)  + \alpha_{1,2} \Li^{(2)}(1,\omega) - \beta_{1,0} \log(1+\omega) - \beta_{1,1} \Li^{(1)}(1,-\omega) \nonumber\\
&+ \beta_{1,2}\Li^{(2)}(1,-\omega) + \beta_{3,0} \Li(3,-\omega) + \beta_{3,2} \Li^{(2)}(3,-\omega) + \beta_{5,0} \Li(5,-\omega) \nonumber\\
&- \beta_{5,1}\Li^{(1)}(5,-\omega) + \beta_{5,2}\Li^{(2)}(5,-\omega) - \beta_{5,3} \Li^{(3)}(5,-\omega) \nonumber\\
&+\sum_{n=1} ^{N^*} \left[\Omega(n) - \Omega^{AS}(n)\right]\omega^n + \mathcal{E}(N^*,\omega) \;
\end{align}
in terms of the polylogarithmic function
\begin{eqnarray}\label{polylog}
    \Li(s,\omega)&=& \frac{-1}{\Gamma(s-1)}\ \int_0^1\frac{dx}{x}\ \log^{s-2}\left(\frac{1}{x}\right)\ \log(1-x\, \omega) \ ,
 \end{eqnarray}
 and its derivatives
 \begin{equation}\label{derivative}
       \Li^{(p)}(s,\omega)\equiv \frac{d^p}{ds^p}\Li(s,\omega)= (-1)^p \sum_{n=1}^\infty \frac{\log^p n}{n^s}\ \omega^n  \ .
 \end{equation}
 Analogous expressions can also be easily obtained for the other 3 channels: it is a simple matter to associate the coefficients $\alpha_{p,k}$ and $\beta_{p,k}$ in Table \ref{table-ab} with the corresponding polylogarithm, after the pattern seen in Eq. (\ref{result3omega}).

\begin{figure}
\begin{center}
\includegraphics[width=7cm] {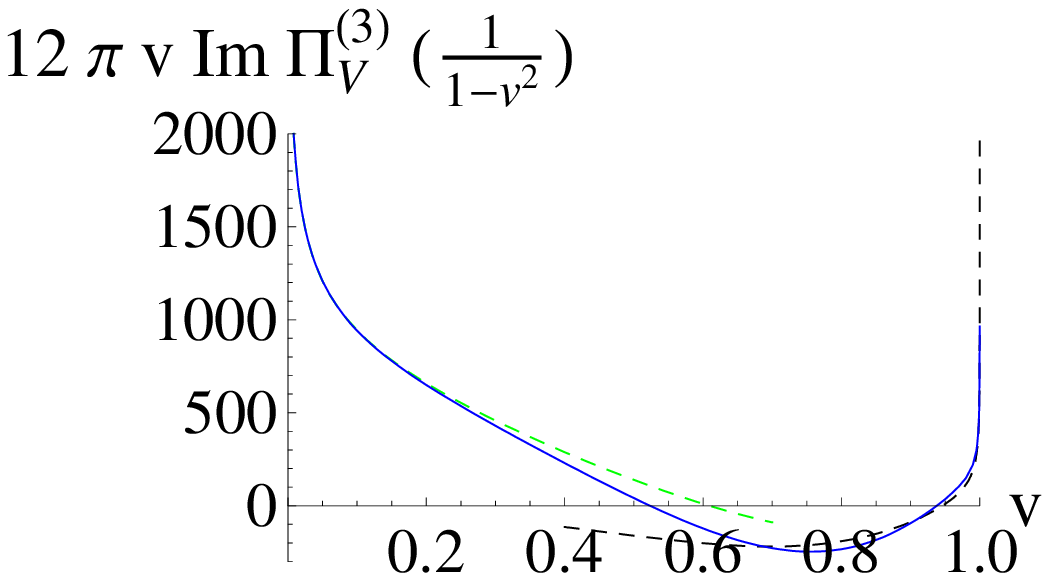}
\includegraphics[width=7cm]{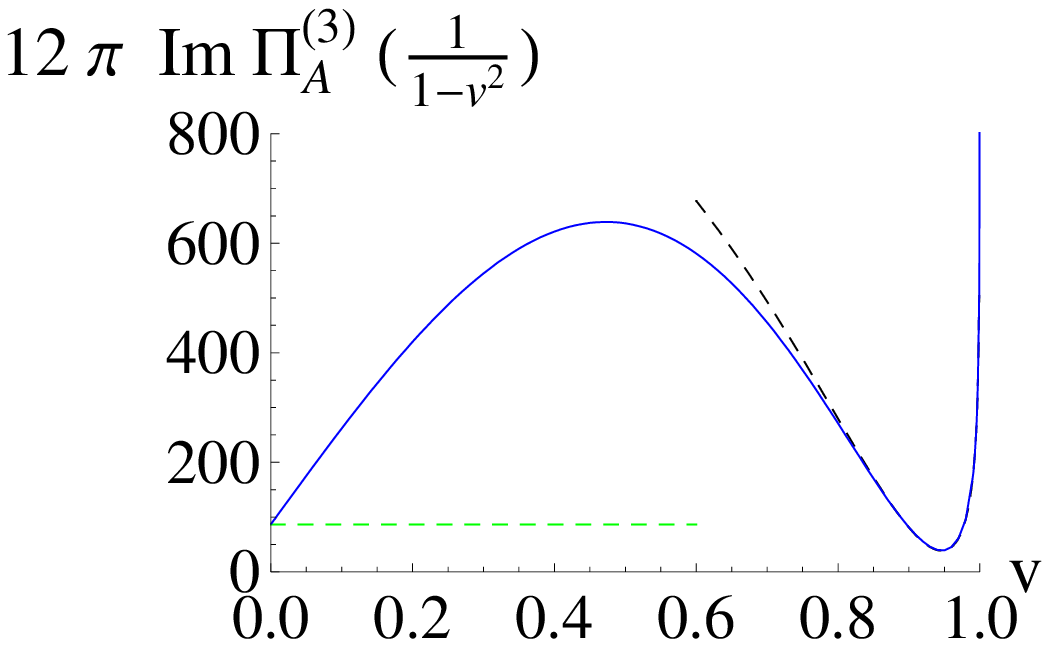}

\vspace{.5cm}

\includegraphics[width=7cm]{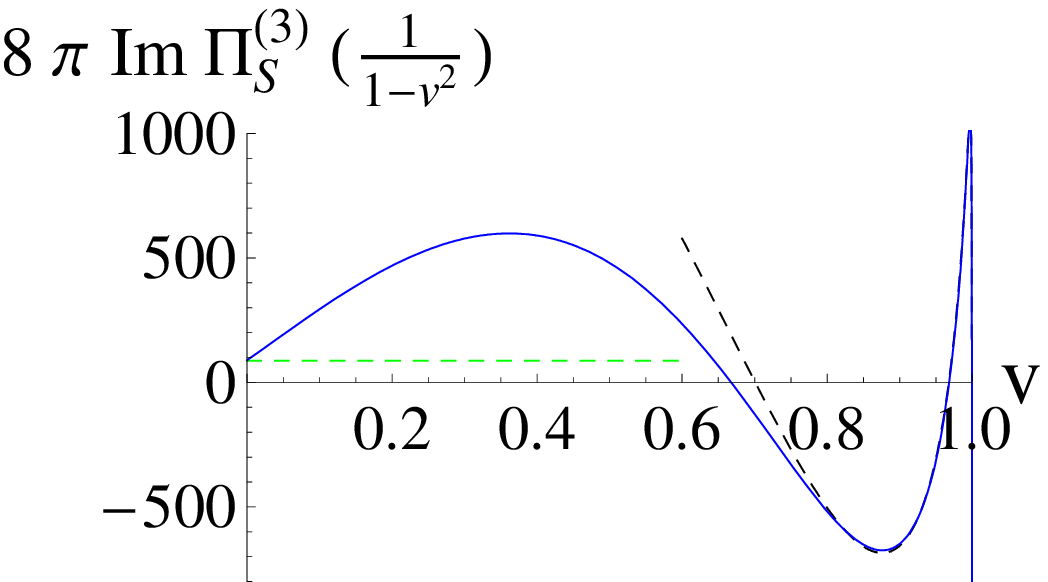}
\includegraphics[width=7cm]{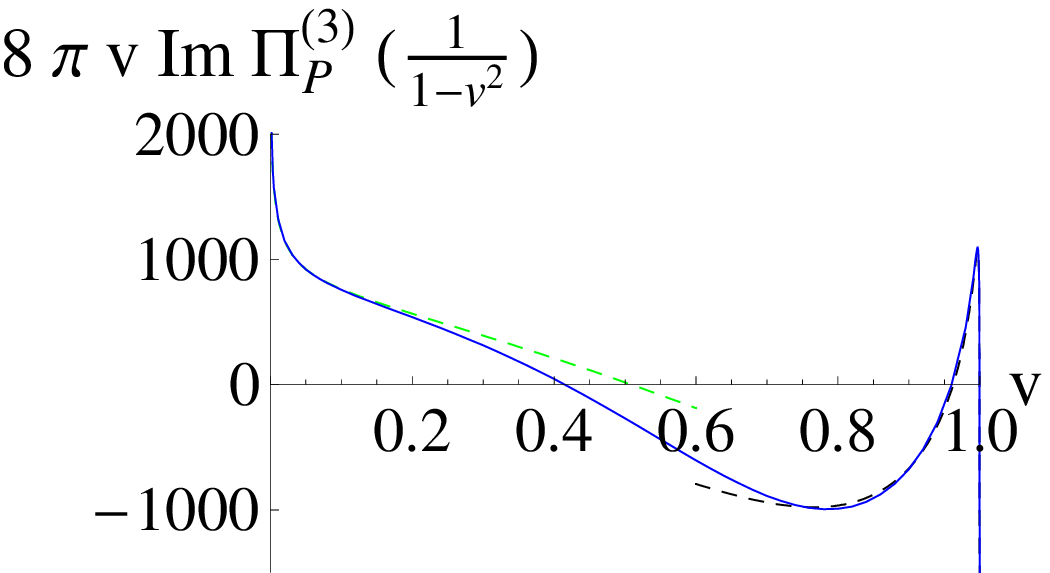}
\end{center}
\caption{Imaginary part of the vector, axial-vector, scalar and pseudo-scalar polarizations as a functions of the velocity $v=\sqrt{1-1/z}$. We also show the threshold expansion (in dashed green) at low $v$, and the OPE (in dashed black) at $v$ near unity. }
\label{fig.spectral_functions}
\end{figure}

For a given number of known Taylor coefficients, $N^*$, the error function (\ref{errorfunction}) may be obtained after a conservative estimate is made of the expected behavior for the difference  $\left[\Omega^{\chi}- \Omega^{\chi}_{AS}\right]$ in Eq. (\ref{errorfunction}). This difference is expected to be of the order of the first unknown term in the asymptotic expansion  (\ref{DefOmega}). Since our present ignorance is dominated by the $\alpha^{\chi}$ coefficients in this expansion, we choose to parameterize this ignorance in the form of an effective $\alpha^{\chi}$. This effective coefficient is shown in Table \ref{totalerrors} for all the channels. Since these effective coefficients affect the value of the non-logarithmic terms of the OPE, i.e. terms of the form $B^{\chi}(p,0)$ in Eq. (\ref{eqB}), we have checked that these estimates are not in conflict with the value of those.

In order to reduce the error even further, one would need to increase the value of $N^*$, i.e. the number of Taylor coefficients in (\ref{eqT}), or the number of nonanalytic OPE coefficients $B^{\chi}(p,k\neq 0)$ in Eq. (\ref{eqB}) but, above all, the most efficient way would be to compute more nonanalytic terms of the threshold expansion, i.e.   $A^{\chi}(p,k\neq 0)$ in Eq. (\ref{eqA}) (see the Appendix for some illustration on this point).

Having an explicit reconstruction of the vacuum polarization function, e.g. (\ref{result3omega}), and an estimate of the error function, e.g. Table \ref{totalerrors}, it is now a simple matter to also reconstruct the spectral functions through, for instance, the following representation:
 \begin{equation}\label{imaginary}
    \mathrm{Im} \Li(s,\mathrm{e}^{i \varphi})=\frac{\sin \varphi}{\Gamma(s)}\ \int_{0}^{1}dx\ \frac{\log^{s-1}\left( \frac{1}{x}\right)}{x^2+1-2\ x\ \cos \varphi} \quad , \quad 0\leq \varphi \leq \pi \ .
\end{equation}
We show in Figs. \ref{fig.spectral_functions} the result for the spectral functions in the four channels, as a function of the quark velocity $v=\sqrt{1-1/z}$. The error on each of these curves is roughly of the order of the width of the curve itself. These results are in rather good qualitative agreement with the spectral functions found with the Pad\'{e} method in Refs. \cite{Hoang:2008qy} and \cite{Kiyo}, as can be judged from the corresponding plots. A more detailed comparison would require knowledge of the error function associated with the Pad\'{e} method, which has become available to us only in the vector channel.\footnote{ A.~H.~Hoang and  V.~Mateu, private communication.}  For instance, at $v=0.75$, we obtain $12 \pi v \mathrm{Im}\Pi^{(3)}_{V}(1/1-v^2)=-246.7(8.8)$, whereas Ref. \cite{Hoang:2008qy} obtains $-219.0(3.4)$,  and Ref. \cite{Kiyo} gets $-214.6$, with an error that we have not been able to quantify in any detail.

Knowledge of the asymptotic expression (\ref{DefOmega}) allows us to predict the Taylor coefficients $C^{\chi}(n)$ through Eq. (\ref{C=Omega0}) by separating the first $N^*$ known terms in this sum and using the asymptotic expression from the term $N^*+1$ onwards. Since these coefficients are of much phenomenological interest for the present determination of the heavy quark masses \cite{Kuhn:2007vp}-\cite{Allison:2008xk} we also list them in Tables \ref{Cvector}-\ref{Cscalar}, together with the analogous results found in Ref. \cite{Kiyo} (see also Ref. \cite{Hoang:2008qy}). Our results confirm those of \cite{Kiyo}, albeit the errors quoted in \cite{Kiyo} are always significantly smaller.

\begin{table}
\centering
{\scriptsize
\begin{tabular}{|c||cc|cc||cc|cc||}
\hline
\multirow{2}{*}{$\frac{16\pi^2}{3}\;C^v(n)$} & \multicolumn{4}{c||}{This work} & \multicolumn{4}{c||}{Ref. \cite{Kiyo}}\\
\cline{2-9}
&\multicolumn{2}{c|}{$n_l=3$} &\multicolumn{2}{c||}{$n_l=4$}&\multicolumn{2}{c|}{$n_l=3$}&\multicolumn{2}{c||}{$n_l=4$}\\
\cline{1-1}
$n$ & & Error &  & Error &  & Error &  & Error\\
\hline\hline
1	&	366.1748&0&308.0188&0&	366.1748&0	&	308.0188&0\\
2	&	381.5091&0&330.5835&0&	381.5091&0	&	330.5835&0\\
3	&	385.2331&0&338.7065&0&	385.2331&0	&	338.7065&0\\
\hline
4	&	382.7&0.5&339.7&0.5&		383.073 &0.011&	339.913&0.010\\
5	&	378.0&1.2&337.7&1.2&		378.688	&0.032&	338.233&0.032\\
6	&	372.5&1.8&334.5&1.8&		373.536	&0.061&	335.320&0.063\\
7	&	367.0&2.3&330.9&2.3	 &		368.23	&0.09	&	331.90 &0.10\\
8	&	361.5&2.7&327.2&2.7&		363.03	&0.13	&	328.33 &0.14\\
9	&	356.4&3.1&323.5&3.1&		358.06	&0.17 &	324.78 &0.18\\
10&	351.6&3.4&320.0&3.4	 &		353.35	&0.20	&	321.31 &0.22\\
\hline
\end{tabular}
}
\caption{Vector channel.}\label{Cvector}
\vspace{.5cm}
\centering
{\scriptsize
\begin{tabular}{|c||cc|cc||cc|cc||}
\hline
\multirow{2}{*}{$\frac{16\pi^2}{3}\;C^a(n)$} & \multicolumn{4}{c||}{This work} & \multicolumn{4}{c||}{Ref. \cite{Kiyo}}\\
\cline{2-9}
&\multicolumn{2}{c|}{$n_l=3$} &\multicolumn{2}{c||}{$n_l=4$}&\multicolumn{2}{c|}{$n_l=3$}&\multicolumn{2}{c||}{$n_l=4$}\\
\cline{1-1}
$n$ & & Error &  & Error &  & Error &  & Error\\
\hline\hline
1&165.1328&0&138.1938&0&165.1328&0&138.1938&0\\
2&105.1185&0&90.0956&0&105.1185&0&90.0956&0\\
3&75.5564&0&65.5198&0&75.5564&0&65.5198&0\\
\hline
4&57.7&0.03&50.42&0.03&57.7298&0.0029&50.4287&0.0042\\
5&46&0.1&40.4&0.1&46.005&0.009&40.397&0.013\\
6&37.8&0.1&33.3&0.1&37.813&0.017&33.338&0.024\\
7&31.8&0.2&28.1&0.2&31.825&0.025&28.151&0.036\\
8&27.3&0.2&24.2&0.2&27.291&0.034&24.206&0.048\\
9&23.7&0.2&21.1&0.2&23.759&0.042&21.123&0.059\\
10&20.9&0.2&18.6&0.2&20.943&0.049&18.658&0.069\\
\hline
\end{tabular}
}
\caption{Axial channel.}\label{Caxial}
\vspace{.5cm}
\centering
{\scriptsize
\begin{tabular}{|c||cc|cc||cc|cc||}
\hline
\multirow{2}{*}{$\frac{16\pi^2}{3}\;C^p(n)$} & \multicolumn{4}{c||}{This work} & \multicolumn{4}{c||}{Ref. \cite{Kiyo}}\\
\cline{2-9}
&\multicolumn{2}{c|}{$n_l=3$} &\multicolumn{2}{c||}{$n_l=4$}&\multicolumn{2}{c|}{$n_l=3$}&\multicolumn{2}{c||}{$n_l=4$}\\
\cline{1-1}
$n$ & & Error &  & Error &  & Error &  & Error\\
\hline\hline
1&16.0615&0&8.6753&0&16.0615&0&8.6753&0\\
2&230.9502&0&199.8289&0&230.9502&0&199.8289&0\\
3&320.5093&0&283.8922&0&320.5093&0&283.8922&0\\
4&359.1116&0&321.5253&0&359.1116&0&321.5253&0\\
\hline
5&376.3&0.3&338.2&0.3&376.3673&0.0023&339.2386&0.0021\\
6&383.4&0.9&345.9&0.9&383.6206&0.0084&347.4338&0.0075\\
7&385.4&1.5&348.8&1.7&385.794&0.018&350.695&0.017\\
8&385.7&2.2&349.1&2.4&385.250&0.032&351.252&0.029\\
9&382.5&2.8&347.9&3.1&383.215&0.048&350.278&0.044\\
10&379.5&3.4&345.9&3.8&380.360&0.066&348.424&0.061\\
\hline
\end{tabular}
}
\caption{Pseudoscalar channel}\label{Cpseudo}
\vspace{.5cm}
\centering
{\scriptsize
\begin{tabular}{|c||cc|cc||cc|cc||}
\hline
\multirow{2}{*}{$\frac{16\pi^2}{3}\;C^s(n)$} & \multicolumn{4}{c||}{This work} & \multicolumn{4}{c||}{Ref. \cite{Kiyo}}\\
\cline{2-9}
&\multicolumn{2}{c|}{$n_l=3$} &\multicolumn{2}{c||}{$n_l=4$}&\multicolumn{2}{c|}{$n_l=3$}&\multicolumn{2}{c||}{$n_l=4$}\\
\cline{1-1}
$n$ & & Error &  & Error &  & Error &  & Error\\
\hline\hline
1&-2.0665 &0&-3.9663 &0&-2.0665&0&-3.9663&0\\
2&59.9301&0&49.7941 &0&59.9301&0&49.7941&0\\
3&69.5687&0&59.9811&0&69.5687&0&59.9811&0\\
\hline
4&64.8&0.3&56.6&0.3&64.641&0.014&56.534&0.014\\
5&57.4&0.8&50.6&0.8&57.168&0.043&50.399&0.041\\
6&50.4&1.2& 44.7&1.2&50.069&0.081&44.374&0.076\\
7&44.4&1.5& 39.5&1.5&43.95&0.12&39.10&0.12\\
8&39.3&1.8& 35.1&1.8&38.81&0.16&34.64&0.15\\
9&35.1&2.1& 31.3&2.1&34.52&0.20&30.89&0.19\\
10&31.5&2.3& 28.2&2.3&30.93&0.24&27.73&0.22\\
\hline
\end{tabular}
}
\caption{Scalar channel}\label{Cscalar}
\end{table}

\section{Comparison with the Pad\'{e} method}\lbl{comparison}
\setcounter{equation}{0}
\def\theequation{\arabic{section}.\arabic{equation}}

By construction, the asymptotic expansion presented in this work in Eq. (\ref{DefOmega}) contains a finite sum which stops at a given order in $k$ and $p$. These indices are determined by the number of \emph{nonanalytic} coefficients, $A(p,k)$ and $B(p,k)$,  which are known from the OPE and threshold expansions (\ref{eqA},\ref{eqB}) \footnote{By nonanalytic coefficients, we mean those $A(p,k)$ and $B(p,k)$, with $p<0, \forall k$ or $p\geq 0, k\neq0$, i.e. those which accompany non-analytic behavior in $1-z$ and $1/z$, respectively.}.  This means, in particular, that there is no way to predict the next term in the OPE or the threshold expansion from the previously calculated terms. This ignorance must be completely encoded in the error function \ref{errorfunction} (see Table \ref{totalerrors}). The situation is similar to what happens in a generic Taylor expansion, where knowledge of $n$ terms does not give any information about the term $n+1$.

This is in contradistinction to the Pad\'{e} method.  In this method, after all the necessary choices are made, one obtains a family of functions which, upon reexpansion around $z\rightarrow -\infty$ and $z\rightarrow 1$, yields a \emph{prediction} for an infinite sequence of coefficients in both the OPE and the threshold expansion. Therefore, this means that the functions reconstructed via the Pad\'{e}-inspired method contain more information than those reconstructed via the method of the present work. This could be consistent with the fact that the errors quoted within the Pad\'{e} method always come out to be smaller, as shown in the Tables \ref{Cvector}-\ref{Cscalar}. An important question to answer, however, is whether the prediction for the OPE and threshold expansion coefficients made by the Pad\'{e} method may be  considered sufficiently accurate. To clearly answer this question, the best will be to compare to the exact result, once the exact result for the next coefficient becomes available using standard diagrammatic techniques.

In order to facilitate this comparison in the future, we use the one example of reconstructed function provided in Ref. \cite{Kiyo} for each channel,  and collect the prediction for the first coefficients in both the OPE and threshold expansion. This we do in both the $z$ and $\omega$ variables in Table \ref{table-Kz} and Table \ref{table-Kw}, respectively. We also deduce, through the matching condition (\ref{MellinPit}), the corresponding $\alpha_{p,k}$ and $\beta_{p,k}$ coefficients in the asymptotic expansion (\ref{DefOmega}), and list them in Table \ref{table-Kab}. Finally, we repeat the same exercise but using the family of reconstructed functions provided by the authors of Ref. \cite{Hoang:2008qy}. In this reference they considered the vector channel only, but they produced a set of Pad\'{e}-reconstructed functions which passed all the good-quality criteria imposed by these authors. The results for this case are listed in Table \ref{table-Hoang}. The two numbers quoted in every entry of this table correspond to the maximum and minimum value obtained within the set of reconstructed functions. When the two numbers are identical, this means that the number predicted turns out to be unique. By comparing the results shown in these tables one can see that sometimes there is complete agreement between the two analyses (like in the case of the $A(1/2,2)$ coefficient), but  other times the agreement looks more  questionable (like in the case of the $B(3,4)$ coefficient). It will be very interesting to compare these predictions to the exact results, once they will become available in the future.

\begin{table}
\centering
{\scriptsize
\begin{tabular}{|c||c|c||c|c||c|c||c|c|}
\hline
& \multicolumn{2}{c||}{$v$} & \multicolumn{2}{c||}{$a$} &\multicolumn{2}{c||}{$p$} & \multicolumn{2}{c|}{$s$}\\
\cline{2-9}
		$\Pi^\chi$ 						&	$n_l=3$ & $n_l=4$ 				&  $n_l=3$ & $n_l=4$ 			&	$n_l=3$ & $n_l=4$ 			 &	$n_l=3$ & $n_l=4$ \\
\hline
\hline
$A(1/2,0)$  & -143.0996& -76.2099& -48.2359  	& -44.0331  & -156.1437	& -161.1861 & -78.7753	 &-70.4228			\\
$A(1/2,1)$  &-13.7341  & -6.4188 & 0  			& 0 		& -15.8236  & -15.6340 & 0			&			 0			\\
$A(1/2,2)$  & 3.2229   & 2.7631  & 0  			& 0 		& 3.2229    & 2.7631   & 0			&			 0			\\
\hline
$B$(3,1)  & -1.2191 & -1.2247  & 0.9647  & 0.8049   & 0.4098    & 0.2578    &  4.3843 	    & 3.8451 \\
$B$(3,2)  &-0.2746  &-0.4585   & 0.1063  &  0.0872  &-1.5633    &-1.4233    &   0.5702      &  0.4616 \\
$B$(3,3)  & 0.2560  & 0.4425   & -0.2087 & -0.1611  & 0.2861    & 0.2187    &	  -0.7012 	& -0.9879  \\
$B$(3,4)  &-0.0073  &-0.0742   &-0.0046  &  0.0049  &   0.0739  &  0.0441   &  -0.0275	    & -0.0068 \\
\hline
\end{tabular}
}
\caption{Predicted coefficients for the OPE and threshold expansion in Eqs. (\ref{eqA},\ref{eqB}) for $n_l=3,4$  using the results of Ref~\cite{Kiyo}.}\label{table-Kz}

  \centering
{\scriptsize
\begin{tabular}{|c||c|c||c|c||c|c||c|c|}
\hline
 & \multicolumn{2}{c||}{$v$} & \multicolumn{2}{c||}{$a$} &\multicolumn{2}{c||}{$p$} & \multicolumn{2}{c|}{$s$}\\
\cline{2-9}
				$\Omega^\chi$				&	$n_l=3$ & $n_l=4$ 				&  $n_l=3$ & $n_l=4$ 			&	 $n_l=3$ & $n_l=4$ 			 &	 $n_l=3$ & $n_l=4$ \\
\hline
\hline
$\Omega^{\chi(+)}(1,0)$  & 37.5996   & 12.0638 & 23.3869& 21.2855 	& 44.7521	& 50.5119  & 38.2910 & 34.1148 \\
$\Omega^{\chi(+)}(1,1)$  & 35.4012   & 25.9895 & 0 		& 0 		& 44.8394   & 42.4421  & 0       & 0	 \\
$\Omega^{\chi(+)}(1,2)$  & -11.5083  & -9.8665 & 0 		& 0 		& -11.5083  &-9.8665   & 0       & 0 \\
\hline
$\Omega^{\chi(-)}(6,1)$  & 0.6588  & 0.4693  & -4.1789 & -3.3437 & -1.4412 & -1.2240& -9.3411  & -7.7181	 \\
$\Omega^{\chi(-)}(6,2)$  & -0.6921 & -0.5036 & -0.1225 & -0.2464 & -0.5123 & -0.4476& -3.6175  & -3.3982	\\
$\Omega^{\chi(-)}(6,3)$  & -0.3031 & -0.3166 & 0.8217  & 0.7098	 &  0.3593 & 0.2710 & 1.3931   & 0.9956	 \\
$\Omega^{\chi(-)}(6,4)$  & -0.2341 & -0.2016 & 0.5560  & 0.5083	 & 0.1412  & 0.1334 & 3.3030   & 3.0905		 \\
\hline
\end{tabular}
}
\caption{Predicted coefficients for the OPE and threshold expansion in the $\omega$-variable (\ref{ThresholdPiw}) for $n_l=3,4$, using the results of Ref~\cite{Kiyo}.}\label{table-Kw}

  \centering
{\scriptsize
\begin{tabular}{|c||c|c||c|c||c|c||c|c|}
\hline
& \multicolumn{2}{c||}{$v$} & \multicolumn{2}{c||}{$a$} &\multicolumn{2}{c||}{$p$} & \multicolumn{2}{c|}{$s$}\\
\cline{2-9}
				$\Omega^\chi_{AS}$ 	&	$n_l=3$ & $n_l=4$ 				&  $n_l=3$ & $n_l=4$ 			&	 $n_l=3$ & $n_l=4$ 			 &	$n_l=3$ & $n_l=4$ \\
\hline
\hline
$\alpha^{\chi}_{2,0}$ & -19.8034 & -11.7994	& 0			& 0 		& -21.8929	& -21.0146	& 0			& 0 			 \\
$\alpha^{\chi}_{2,1}$ & -12.8916 & -11.0524 & 0			& 0 		& -12.8916  & -11.0524	& 0			& 0				 \\
$\alpha^{\chi}_{2,2}$ & 0        &  0       & 0			& 0 		& 0       	& 0         & 0			& 0\\
\hline
$\beta^{\chi}_{7,0}$ & -282.1379 & 116.5165 & 363.6681  & 324.5732  & -1150.5165& -1030.2552& 2128.4420	& 1646.6387	\\
$\beta^{\chi}_{7,1}$ & 238.9509  &-120.5650	& -307.8491 &-228.3280  & 846.5914	& 632.6807	& -2060.9357& -1931.8120\\
$\beta^{\chi}_{7,2}$ & 82.7196   &-50.8681	& -112.0433 &-100.2002  & 372.6561  & 334.8105  & -660.5808 & -499.3996 \\
$\beta^{\chi}_{7,3}$ & -32.4972  & 18.2039  & 43.6160   & 34.0066   & -123.2945 & -97.8701  & 260.2617  & 231.4261  \\
\hline
\end{tabular}
}
\caption{Induced values for the coefficients in the asymptotic expansion (\ref{DefOmega}) for $n_l=3,4$  corresponding to the predicted terms in both the OPE and threshold expansion listed in Tables \ref{table-Kz} and \ref{table-Kw}.}\label{table-Kab}

  \centering
{\scriptsize
\begin{tabular}{|c||c|c||c||c|c||c||c|c|}
\hline
 & \multicolumn{2}{c||}{$v$ } & & \multicolumn{2}{c||}{$v$ } &  &\multicolumn{2}{c|}{$v$ } \\
\cline{2-3}
\cline{5-6}
\cline{8-9}
	$\Pi^v$ 							&	$n_l=3$ & $n_l=4$ 	&$\Omega^v$ 	&  $n_l=3$ & $n_l=4$ & $\Omega^v_{AS}$	&	$n_l=3$ & $n_l=4$  \\
\hline
\hline
$A(1/2,0)$ &$\begin{matrix} -49.6008\\ -291.5004 \end{matrix}$ & $\begin{matrix} -37.7892\\ -106.0299 \end{matrix}$ & $\Omega^{(+)}(1,0)$ &  $\begin{matrix}111.8000 \\ 1.5961 \end{matrix}$      & $\begin{matrix}22.0766 \\-2.0938\end{matrix}$&  $\alpha_{2,0}$ &$\begin{matrix}-4.2693\\-19.8693\end{matrix}$ & $\begin{matrix}-4.4806\\-18.8806 \end{matrix}$\\
\cline{2-3}
\cline{5-6}
\cline{8-9}
$A(1/2,1)$ & $\begin{matrix} 1.7691 \\ -13.7341 \end{matrix}$  &$\begin{matrix}0.8707\\ -13.4841\end{matrix}$& $\Omega^{(+)}(1,1)$ & $\begin{matrix}35.4012 \\ 19.8981 \end{matrix}$     & $\begin{matrix}33.0547\\18.6999\end{matrix}$ &  $\alpha_{2,1}$ &$\begin{matrix}-12.8916\\-12.8916\end{matrix}$ &$\begin{matrix}-11.0524\\-11.0524\end{matrix}$ \\
\cline{2-3}
\cline{5-6}
\cline{8-9}
$A(1/2,2)$ & $\begin{matrix}3.2229 \\ 3.2229 \end{matrix}$  &$\begin{matrix}2.7631\\2.7631\end{matrix}$ & $\Omega^{(+)}(1,2)$ &$\begin{matrix} -11.5083 \\  -11.5083 \end{matrix}$ &$\begin{matrix} -9.8665 \\   -9.8665 \end{matrix}$&  $\alpha_{2,2}$ & 0          & 0          \\
\cline{2-3}
\cline{5-6}
\cline{8-9}
\hline
$B$(3,1) &$\begin{matrix}-0.9944 \\-1.2222  \end{matrix}$ &$\begin{matrix}-1.1203\\-1.3309\end{matrix}$& $\Omega^{(-)}(6,1)$ &  $\begin{matrix} 1.3333 \\ 0.7307 \end{matrix}$  & $\begin{matrix}1.1538 \\0.6148\end{matrix}$&  $\beta_{7,0}$ & $\begin{matrix}547.6502\\-237.6193 \end{matrix}$  & $\begin{matrix}497.1674\\-202.8925 \end{matrix}$          \\
\cline{2-3}
\cline{5-6}
\cline{8-9}
$B$(3,2) &$\begin{matrix}0.8006 \\ 0.2451 \end{matrix}$&$\begin{matrix}0.6936\\0.1830\end{matrix}$& $\Omega^{(-)}(6,2)$ &$\begin{matrix} -0.7378 \\-1.2126 \end{matrix}$  & $\begin{matrix}-0.6553\\-1.0802 \end{matrix}$  &  $\beta_{7,1}$ &$\begin{matrix}585.1540\\224.1701\end{matrix}$  & $\begin{matrix}541.7054\\220.6132 \end{matrix}$     \\
\cline{2-3}
\cline{5-6}
\cline{8-9}
$B$(3,3) & $\begin{matrix}1.2853 \\ 0.3659 \end{matrix}$    & $\begin{matrix}1.1730\\0.3462 \end{matrix}$ & $\Omega^{(-)}(6,3)$ &  $\begin{matrix}-0.1656 \\-0.2957\end{matrix}$  & $\begin{matrix}0.2324\\-0.1236\end{matrix} $&  $\beta_{7,2}$ & $\begin{matrix}60.6614\\-187.7386\end{matrix}$& $\begin{matrix}62.7959\\-168.4818 \end{matrix}$        \\
\cline{2-3}
\cline{5-6}
\cline{8-9}
$B$(3,4) &$\begin{matrix}0.0119 \\ 0.0010 \end{matrix}$  &$\begin{matrix}0.0102\\0.0008\end{matrix}$ &$\Omega^{(-)}(6,4)$ &$\begin{matrix} -0.2357\\-0.2330\end{matrix}$&$\begin{matrix}-0.2176\\-0.2199 \end{matrix}$&  $\beta_{7,3}$ &$\begin{matrix}-38.4732\\-46.3932\end{matrix}$      & $\begin{matrix}-42.5641\\-45.0972 \end{matrix}$        \\
\cline{2-3}
\cline{5-6}
\cline{8-9}
\hline
\end{tabular}
}
\caption{Predicted coefficients corresponding to the first unknown term in the OPE and threshold expansion in the $z$-variable (second and third columns) and in the $\omega$-variable (fifth and sixth columns) from the results of Ref~\cite{Hoang:2008qy}. The corresponding coefficients in the asymptotic expansion (\ref{DefOmega}) are also deduced (eight and ninth columns). See text.}\label{table-Hoang}
\end{table}

\section{Summary and Conclusions}\lbl{conclusions}
\setcounter{equation}{0}
\def\theequation{\arabic{section}.\arabic{equation}}

In this work we have applied a method, previously introduced in Ref. \cite{Greynat}, to reconstruct all the  vacuum polarization functions (\ref{poldef}) in the whole complex plane. The method uses the first coefficients from the Taylor expansion (\ref{eqT}) and the nonanalytic coefficients of the OPE and threshold expansions (\ref{eqA},\ref{eqB}) to resum these three expansions simultaneously. The reconstruction is carried out by means of a systematic calculation of the asymptotic coefficients $\Omega(n)$ (\ref{DefOmega}) for large $n$, which control the expansion of the vacuum polarization functions in the conformal variable $\omega$ (\ref{conformal}). This reconstruction consists of an approximation to the original function in terms of a finite number of polylogarithms (and their derivatives) plus a well-defined polynomial, as shown in Eq. (\ref{result3omega}). The approximation is controlled by an error function (\ref{errorfunction}) which becomes smaller as more information is known.  In this regard, we emphasize that our present errors are dominated by the unknown terms in the threshold expansion. Therefore, a calculation of the coefficients $A(1/2,k\neq 0)$ in the expansion  (\ref{eqA}) would significantly improve on the precision achieved in the present determination. Alternatively, knowledge of more terms in the Taylor expansion (\ref{eqT}) (i.e. a larger $N^*$) would also help, since currently $N^*=3,4$, only. Given that at $\mathcal{O}(\alpha_s^2)$ nothing less than 30 terms are known in the vector channel \cite{BCS,MMM}, we may optimistically look forward to an improvement on this in the future.

\vspace{1cm}

\textbf{Acknowledgements}

\vspace{.2cm}
\noindent We thank  Y. Kiyo, A. Maier, P. Maierh\"{o}fer and P. Marquard for correspondence; and A. Hoang and V. Mateu for discussions and for making the files with their results available to us.
This work has been supported by MICINN (grant FPA2009-09638), CICYT-FEDER-FPA2008-01430, SGR2005-00916, the Spanish Consolider-Ingenio 2010 Program CPAN (CSD2007-00042), and by Junta de Andaluc\'{i}a (Grants P07-FQM 03048 and P08-FQM 101).

\newpage

\textbf{\Large APPENDIX }

\begin{figure}
\begin{center}
\includegraphics[width=3.5in]{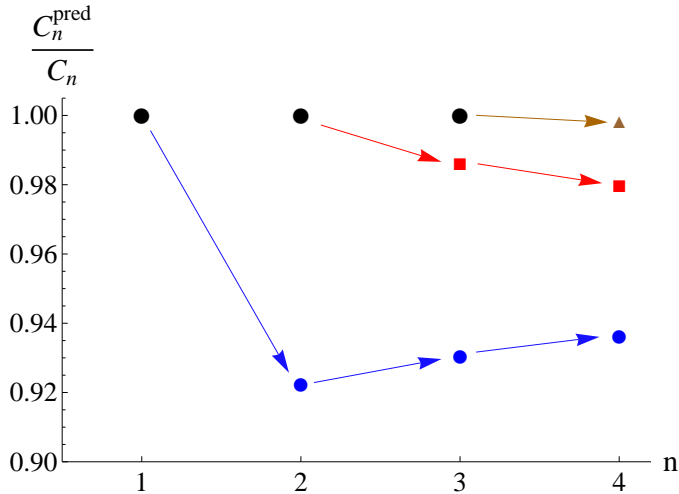}
\end{center}
\caption{Comparison between the prediction for the next Taylor coefficients in the pseudoscalar channel and the exact result, as a function of the number of Taylor coefficients known, $N^*$. See text. }
\label{fig.CnPseudoscalar}
\end{figure}

\vspace{1cm}

The pseudoscalar channel allows us to check the convergence properties of our approximation in the following way. As we did at the end of section 3, we can compute the coefficients $C^{p}(n)$ using Eq. (\ref{C=Omega0}) by separating the first $N^*$ terms in this sum, which are exactly known, and using the asymptotic expression (\ref{DefOmega}) from the term $N^*+1$ onwards. By pretending that only one term is known, i.e. $N^*=1$, one can make a prediction for the other coefficients $C^{p}(n)$ with $n=2,3,4$ and then compare to their exact value. The result of this exercise is shown in Fig.  \ref{fig.CnPseudoscalar} where the black dot signifies the exact value known for $C^{p}(1)$ and the arrow leading to the blue dots are the predictions for  the coefficients $C^{p}(2)-C^{p}(4)$ obtained in this way. Alternatively, one may now pretend that two coefficients are known, i.e. $N^*=2$, and then predict the other two coefficients: this result is represented by the red squares. Finally, if three coefficients are assumed to be known, i.e. $N^*=3$, the prediction for the fourth term is the brown triangle. As one can see, there is a very nice convergence as $N^*$ is increased, already for very low values of $N^*$.

\begin{figure}
\begin{center}
\includegraphics[width=3in]{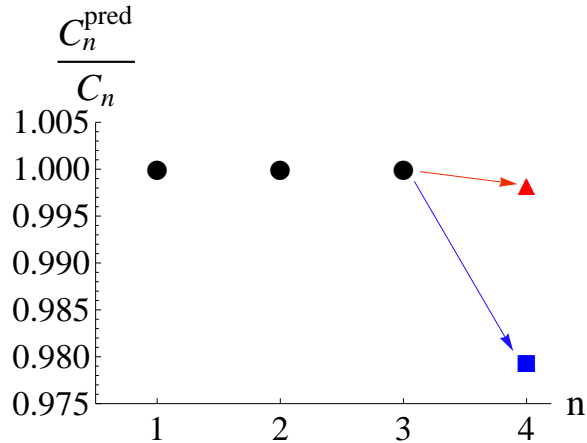}
\end{center}
\caption{Comparison between the prediction for $C^p(4)$ and the exact result, as a function of the number of terms kept in the asymptotic expansion (\ref{DefOmega}). See text.}
\label{fig.C4Pseudoscalar}
\end{figure}

The previous exercise was done with the asymptotic expression (\ref{DefOmega}) including all the coefficients $\alpha^p$ and $\beta^p$ obtained in Table \ref{table-ab} for the pseudoscalar channel. Taking $C^{p}(1)-C^{p}(3)$ as exactly known, we may also test the convergence properties of the asymptotic expansion (\ref{DefOmega}) by comparing the result for the coefficient $C^p(4)$ and the exact result if we now pretend that the last coefficients $\alpha^{p}_{1,k}, (k=0,1,2)$ and $\beta^{p}_{5,k}, (k=0,1,2,3)$ are not known.\footnote{The main effect is due to the threshold coefficients $\alpha^{p}$. The OPE coefficients $\beta^{p}$ play a minor role.} The result is shown in Fig. \ref{fig.C4Pseudoscalar} as a blue square. This should be compared to the red triangle, which is the prediction obtained when all the coefficients  $\alpha^p$ and $\beta^p$ in Table \ref{table-ab} are included. Again, this shows that the asymptotic expression (\ref{DefOmega}) converges nicely to the right result.

\end{document}